\documentclass{sig-alt-release2}
\usepackage[tight,footnotesize]{subfigure}
\usepackage{cite}

\begin{document}
\conferenceinfo{WWW 2012 Companion,} {April 16--20, 2012, Lyon, France.} 
\CopyrightYear{2012} 
\crdata{978-1-4503-1230-1/12/04} 
\clubpenalty=10000 
\widowpenalty = 10000
\title{Multiple Spreaders Affect the Indirect Influence on Twitter}
%
%
%
%
%

\numberofauthors{6} 
%
\author{
%
%
\alignauthor
Xin Shuai\\
       \affaddr{School of Informatics and Computing}\\
       \affaddr{Indiana University Bloomington}\\
       \affaddr{IN, USA}\\
       \email{xshuai@indiana.edu}
\alignauthor
Ying Ding\\
       \affaddr{School of Library and Information Science}\\
       \affaddr{Indiana University Bloomington}\\
       \affaddr{IN, USA}\\
       \email{dingying@indiana.edu}
\alignauthor 
Jerome Busemeyer\\
       \affaddr{Dept. of Psychological and Brain Science}\\
       \affaddr{ Indiana University Bloomington}\\
       \affaddr{IN, USA}\\
       \email{jbusemey@indiana.edu}
}

\maketitle
\begin{abstract}
 Most studies on social influence have focused on direct influence, while another interesting question can be raised as \emph{whether indirect influence exists between two users who're not directly connected in the network and what affects such influence.} In addition, the theory of \emph{complex contagion} tells us that more spreaders will enhance the indirect influence between two users. Our observation of intensity of indirect influence, propagated by $n$ parallel spreaders and quantified by retweeting probability on Twitter , shows that complex contagion is validated globally but is violated locally. In other words, the retweeting probability increases non-monotonically with some local drops. 
\end{abstract}




\section{Introduction}
Social influence has been studied by many researchers. However, most relevant studies focused on direct influence~\cite{xiang, tang} while few touched indirect influence\cite{ying}. Normally, multiple intermediate persons called spreaders are involved in the indirect communication between two persons, i.e., the sender and the receiver. Those spreaders may have a combinational effect on the indirect influence propagated from the sender to the receiver. 




A concept closely related to parallel indirect influence is \emph{complex contagion}, a phenomenon where repeated exposures of an individual to an idea recommended by his/her multiple neighbors positively affect the probability he/she will eventually follow that idea
~\cite{damon}. Romero et al.~\cite{romero} studied the spread of hashtags in Twitter and quantified the probability of a user adopting a new hashtag as the function of the number of his/her neighbors who have already adopted it. They found that the spread of political hashtags validates the complex contagion, where the adoption probability increases monotonically as the number of neighbors who have already adopted the same hashtags increases, until a plateau is finally reached. 

\begin{figure}[htbp]\centering
	\includegraphics[width=\columnwidth]{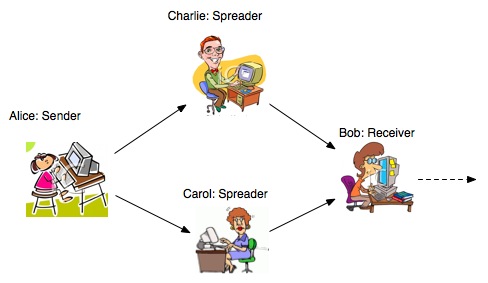}  
	\caption{Typical information spread in a social network}
	\label{fig:flow} 
\end{figure}

The problem we are studying is similar to~\cite{romero}, but we focus on message spread behavior and indirect influence on Twitter. 
A concrete example of this is shown in Figure~\ref{fig:flow},  
where Alice sends out original messages, Charlie and Carol further spread Alice's messages (i.e., by retweeting) and Bob finally receives them. After that, Bob may choose to further spread Alice's messages to others, just like his two neighbors Charlie and Carol have done, or not. 
Here, the intent of Bob to further spread Alice's messages would reflect the intensity of the indirect influence of Alice on Bob, which can be measured as the probability that Bob will further spread Alice's messages, given that Charlie and Carol have already spread these messages. If complex contagion takes effect, the influence intensity will be higher when both Charlie and Carol spread Alice's messages than when either or none of the two spread them. 

In this paper, we examine the intensity of indirect influence as the function of the number of parallel spreaders, between two users on Twitter, who don't have direct following relations. We found that \emph{complex contagion} is observed globally but is violated locally. Especially, when the number of spreaders increases from one to two, there's an obvious drop in the intensity. 
 

\section{Problem Definition}
Twitter user can read another user's messages by following them and re-send their messages via retweeting. 
A retweeting message starts with the identifier ``\emph{RT @username}''. Given a collection of tweets $C=\{t\}$, $V$ represens all Twitter users while $E=\{(u,v)\ |u,v\in V\}$ represents all following relations where $u$ follows $v$. We provide several formal definitions as follows:
\begin{itemize}
\item DEFINITION 1. [Following Triple] $\forall t$ starting with ``\emph{RT @y: RT @x}'' posted by \emph{z}, we build a following triple $T_{xyz} = (x, y, z), x, y, z \in V$ and claim that $(z,y) \in E$ and $(y,x) \in E$. We also define $C(T_{xyz})$ as the total count of tweets that belongs to $T_{xyz}$ and $C(v),v\in V$ as the total count of tweets $v$ posted.
\item DEFINITION 2. [Spreaders] $\forall a,b\in V$, we define spreaders between $a$ and $b$ as $S_{ab} = \{y|T_{ayb}\neq NULL,\\
y\in V\}$.
\item DEFINITION 3. [N-spreader Retweeting Pattern] $\forall a,b\\\in V$ we define a retweeting pattern $P_{ab}=\{ T_{ayb}|y\in S_{ab} \}$ and $|S_{ab}|=n$. Consequently, we define a n-spreader retweeting pattern as $P_n=\{P_{ab}||S_{ab}|=n \}$, and $P_{ab}$ is an $instance$ of $P_n$.
\item DEFINITION 4. [Retweeting Probability] $\forall P_{ab}\neq \varnothing$, we define the probability of $b$ retweeting from $a$ as $\\
Pr(b|a;S_{ab})=\sum_{y\in S_{ab}} C(T_{ayb})/C(a)$. Consequently, we define the retweeting probability of n-spreader retweeting pattern as $Pr(n)=\sum_{P_{ab}\in P_n}Pr\{b|a;S_{ab}\}/|P_n|$. 
\item DEFINITION 5. [Indirect Influence] $\forall x,z \in V \cap P_{xz}\neq \varnothing$, we think $x$ exerts indirect influence on $z$. $Pr(n)$ indicates the average intensity of indirect influence in n-spreader retweeting pattern.
\end{itemize}
 Our research question can be formulated as: \emph{Given $n$ spreaders, how does the curve $Pr(n)$ change with $n$}?
\section{Discussion and Conclusion}
%
The dataset\footnote{http://snap.stanford.edu/data/twitter7.html} contains 467 million tweets from 20 million Twitter users from June to December 2009 , which covers 20\%-30\% of total public tweets during this period. 
\begin{figure}[htbp]\centering
	\includegraphics[width=\columnwidth]{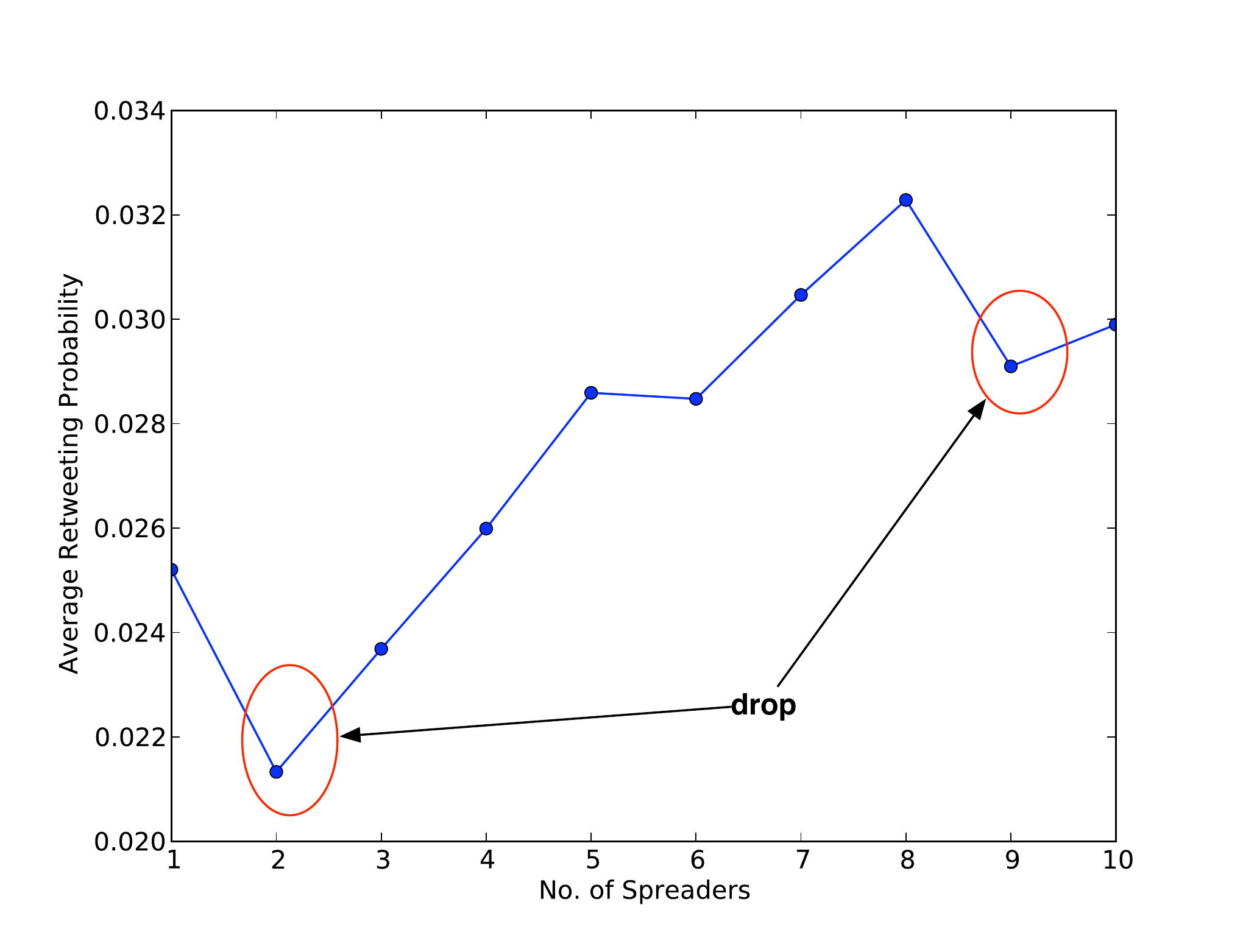}  
	\caption{The curve of $Pr(n)$}
	\label{fig:jure} 
\end{figure}
Figure~\ref{fig:jure} shows the global trend of $Pr(n)$ is increasing as $n$ increases. That is to say, overall, the intensity of indirect influence tends to become higher, or at least persists, as more spreaders are included, which validates the phenomenon of complex contagion in the global level. However, there are two drops spotted in $Pr(n)$, i.e. from $n$ =1 to 2 and 8 to 9. 
We use t-test of \emph{difference between two means} to examine the two hypothesis: $Pr(1) > Pr(2)$ and $Pr(8) > Pr(9)$. Both p-values turn out to be close to zero, 
%
%
implying that the decrease is statistically significant and complex contagion is violated locally.

The emerging field of \emph{quantum cognition} might be able to provide a potential interpretation for the decreased influence phenomenon. Notably, in the process of decision making where a decision depends on multiple factors, quantum cognition assumes that these factors are not independent but have quantum-like interference effects on the final decision in a manner similar to the explanation for results from double-slit experiments~\cite{yukalov}. 
In Figure~\ref{fig:flow}, we assume that initially only Charlie spreads Alice's messages while Carol does not. Bob receives Alice's messages through Charlie and further spreads them because Bob thinks Alice's messages are relevant. Later on, Carol also begins to spread Alice's messages but they largely overlap with those already spread by Charlie. Bob therefore becomes less interested in Alice's messages because he's overwhelmed with redundant information. Thus Bob's intent to further spread Alice's messages (i.e., the indirect influence of Alice on Bob) decreases. Here, the interference between two spreaders leads to destructive effects on the indirect influence from the sender to the receiver. 
In conclusion, we investigated the propagation of parallel indirect influence on Twitter with a focus on how the intensity of influence changes with the number of spreaders. We quantified the intensity of indirect influence with the retweeting probability, and plotted the curve of retweeting probability against the number of spreads. We found that the phenomenon of complex contagion is validated globally but violated locally since the retweeting probability increases non-monotonically with some local drops. We finally proposed quantum cognition hypothesis in an attempt to interpret the local anomaly yet further verification is needed.   

\bibliographystyle{abbrv}
\bibliography{refs}  

\begin{thebibliography}{1}

\bibitem{damon}
D.~Centola and M.~Macy.
\newblock Complex contagion and the weakness of long ties.
\newblock Technical report, 2005.

\bibitem{ying}
Y.~Ding.
\newblock Scientific collaboration and endorsement: Network analysis of
  coauthorship and citation networks.
\newblock {\em Journal of Informetrics}, 5:187--203, 2011.

\bibitem{romero}
D.~M. Romero, B.~Meeder, and J.~Kleinberg.
\newblock Differences in the mechanics of information diffusion across topics:
  idioms, political hashtags, and complex contagion on twitter.
\newblock In {\em WWW '11}, pages 695--704, Hyderabad, India, 2011.

\bibitem{tang}
J.~Tang, J.~Sun, C.~Wang, and Z.~Yang.
\newblock Social influence analysis in large-scale networks.
\newblock In {\em KDD '09}, pages 807--816, Paris, France, 2009.

\bibitem{xiang}
R.~Xiang, J.~Neville, and M.~Rogati.
\newblock Modeling relationship strength in online social networks.
\newblock In {\em WWW '10}, pages 981--990, Hong Kong, China, 2010.

\bibitem{yukalov}
V.~I. Yukalov and D.~Sornette.
\newblock Decision theory with prospect interference and entanglement.
\newblock {\em Theory and Decision}, 70:283--328, 2011.

\end{thebibliography}
\end{document}